\documentclass[twoside,12pt, letterpaper,reqno]{amsart}
\usepackage[dvips]{graphics}
\usepackage[justification=centering,skip=2pt,font=footnotesize]{caption}
\usepackage{amscd,amsfonts,amsmath,amssymb,amsthm,booktabs,color,enumitem,epsfig,fancyhdr,float,graphicx,hyperref,latexsym,marvosym,mathrsfs,mathtools,pstricks,subcaption,tikz,times,url,xstring}
\usepackage[T1]{fontenc}
\usepackage[myheadings]{fullpage}
\usepackage[nosort]{cite}
\usepackage{flowchart}\usetikzlibrary{shapes,arrows,positioning,calc,fit}

\hypersetup{
    pdfborder = {0 0 0}
}
\numberwithin{equation}{section}
\graphicspath{ {./Figures/} }

\newcommand{\doublerule}[1][.4pt]{%
    \noindent
    \makebox[0pt][l]{\rule[.7ex]{\linewidth}{#1}}%
    \rule[.3ex]{\linewidth}{#1}}

\begin{document}

\doublerule[1pt]
\begin{center}
\vspace{-.8cm}
{\Large{\textsc{{Potential of the Sterile Insect Technique for\\[5pt]Control of Deer Ticks, \textit{Ixodes scapularis}}}}}\\
\vspace{.35cm}
\doublerule[1pt]\\
\medskip
by\\\textsc{Thomas Kirby}\\
\smallskip
\textsc{Julie Blackwood, Advisor}\\
\begin{figure}[h!]
\centering
\includegraphics[scale = 0.15]{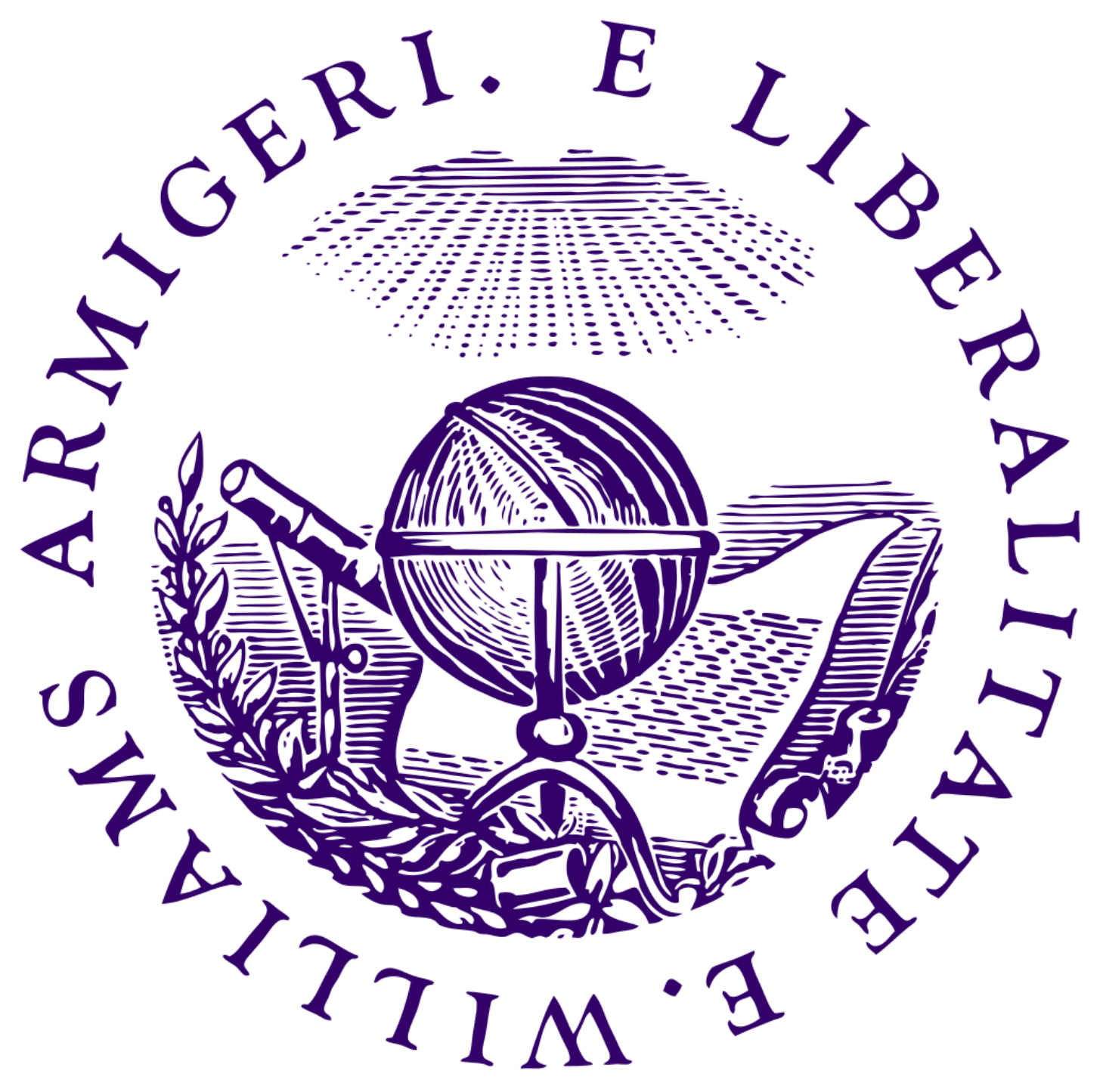}
\label{seal:Williams}
\end{figure}
A thesis submitted in partial fulfillment\\ of the requirements for the
\\Degree of Bachelor of Arts with Honors\\ in Mathematics\\
\smallskip
\textsc{Williams College}
\\
Williamstown, Massachusetts\\
\smallskip
Last revised September 13, 2021
\thispagestyle{empty}
\end{center}
\pagebreak
\pagestyle{plain}
\centerline{\textsc{{Abstract}}}

The deer tick, \textit{Ixodes scapularis}, is a vector for numerous human diseases, including Lyme disease, anaplasmosis, and babesiosis. Concern is rising in the US and abroad as the population and range of this species grow and new diseases emerge. Herein I consider the potential for control of \textit{I. scapularis} using the Sterile Insect Technique (SIT), which acts by reducing net fertility through release of sterile males. I construct a population model with density-dependent and -independent growth, migration, and an Allee effect (decline of the population when it is small), and use this model to simulate sterile tick release in both single- and multi-patch frameworks. I test two key concerns with implementing \textit{I. scapularis} SIT: that the ticks' lengthy life course could make control take too long and that low migration might mean sterile males need thorough manual dispersal to all parts of the control area. Results suggest that typical $\textit{I. scapularis}$ SIT programs will take about eight years, a prediction near the normal range for the technique, but that thorough distribution of sterile ticks over the control area is indeed critical, increasing expense substantially by necessitating aerial release. With particularly high rearing costs also expected for $\textit{I. scapularis}$, the latter finding suggests that cost-effectiveness improvements to aerial release may be a prerequisite to $\textit{I. scapularis}$ SIT.

\pagebreak
\centerline{\textsc{Acknowledgements}}

I extend special thanks to Professor Julie Blackwood, my advisor, for guiding me through the process of developing this work. I am also appreciative of Professor Stewart Johnson and my father, Kris Kirby, for their feedback as readers; my family for their support over the last year; and everyone who attended my thesis defense. Finally, I would like to thank the library staff who hand-scanned the many reference papers I could not access on my own.

\pagebreak
\tableofcontents
\setcounter{tocdepth}{5}
\pagebreak

\section{Introduction}

\subsection{Background}

Ticks are considered the most important vectors of human diseases in North America; worldwide, they are second only to mosquitoes \cite{parola2001ticks,de2016strategies,troughton2007life}. The deer tick or blacklegged tick \textit{Ixodes scapularis} is a vector for numerous serious diseases, including Lyme, anaplasmosis, and babesiosis \cite{troughton2007life}. Population growth and expansion of its range within the northeastern US as well as outward to new regions, driven in large part by reforestation and often accompanied by the spread of pathogens, is increasing concern about the future of \textit{I. scapularis}-borne illness in the US and abroad \cite{ogden2013changing,tran2021spatio,nielsen2020first,burrows2021multi,wu2016impact}. Closely related species and disease vectors in the American southeast and Europe, \textit{Ixodes affinis} and \textit{Ixodes ricinus} (respectively), are also on the rise \cite{nadolny2018modelling,medlock2013driving}; indeed, the genus \textit{Ixodes} could be considered a ``global menace'' \cite{murgia2019meeting}. At the same time, new tick-borne diseases continue to emerge \cite{madison2020emerging}. At present, \textit{I. scapularis} control relies primarily on pesticides \cite{stafford2017integrated,george2000present,stafford1997pesticide}. Unfortunately, pesticide use is replete with drawbacks and limitations, including damage to non-target species and the uncertainty of long-term reliability as tick populations develop resistance \cite{george2000present,de2015tick}.

Herein I model the Sterile Insect Technique (SIT) for \textit{I. scapularis}\footnote{Here and throughout I use the term SIT for simplicity, although in the case of ticks it might more properly be referred to as the Sterile Acarine Technique, or SAT \cite{de2006autocidal}.}. This control method is as yet untested for ticks, but could serve as an alternative to pesticides for smaller populations, for instance as the last stage of an Area-Wide Integrated Pest Management (AW-IPM) program \cite{kiss2012tick,de2006autocidal,klassen2021area,stafford2017integrated}. Although SIT has its own collection of technical, ethical, and ecological concerns \cite{alphey2010sterile,parker2021mass}, these must be weighed against the drawbacks of existing control strategies.

Some scientists consider the Sterile Insect Technique to be the most original idea of the 20\textsuperscript{th} century \cite{speccoll2006edward}. The premise is simple: sterile males are reared and released to compete for mates with wild males, reducing net fertility; with sufficient release the population declines \cite{knipling1955possibilities}. E.F. Knipling conceived the approach in 1937 as a control tactic for the screwworm fly, and the first SIT program was commenced in the 1950s using screwworm flies sterilized by irradiation \cite{lindquist1955use,knipling1959sterile}. By the 1970s, release of sterile screwworm flies was in the hundreds of millions per week, and the northern boundary of the screwworm fly population was eventually pushed from the mid-US down to Panama \cite{novy1991screwworm}. A variety of insects have since been controlled using SIT \cite{klassen2021history,simmons2021impact,hendrichs1983six,vreysen2000glossina}, including, recently, mosquitoes \cite{klassen2021history,benedict2003first,benedict2021sterile}. However, SIT of some species has been determined cost-ineffective, and it remains to be seen how an \textit{I. scapularis} program might fare \cite{hardee2003eradicating,simmons2021impact,klassen2021history}.

Use of SIT for ticks was first proposed in 1982, with a basic preliminary model predicting success \cite{osburn1982potential}. At the time, sterile tick production was achieved by hybridization of two species of cattle tick. Recent progress in genetic engineering and the sequencing of the first tick genome, conveniently that of \textit{I. scapularis} \cite{gulia2016genomic}, have created renewed interest in tick SIT \cite{de2006autocidal,de2016tick}. In 2006, it was discovered that RNA interference targeting the Subolesin gene produces sterility in male ticks, including \textit{I. scapularis} \cite{de2006autocidal,kocan2007transovarial,kocan2015insights}. A proof-of-concept test on cattle ticks also conveniently found a dramatic reduction in the ability of the modified ticks to transmit pathogens \cite{merino2011targeting}. Attempts at multi-generation knockdown by injection of replete females have met with limited success \cite{merino2011control}, but continued developments in gene editing such as CRISPR and the ReMOT Control method hold promise for the future of \textit{I. scapularis} sterilization \cite{chaverra2018targeted,de2018controlling}.

\subsection{Modeling and objectives}

Despite being mentioned with some regularity in the tick literature, tick SIT has hardly been modeled. A few old models simulate SIT for cattle ticks, but these models have little bearing on modern \textit{I. scapularis} SIT \cite{osburn1982potential,weidhaas1983basic,lodos1999simulation}. Herein I develop a deterministic, discrete-time population model for \textit{I. scapularis} SIT.

Two traits of \textit{I. scapularis} set them well apart from previous SIT targets: a much longer life cycle and relative immobility \cite{yuval1990duration,lance2021mass,falco1991horizontal,carroll1996dispersal,troughton2007life,vargas2021impact,zethner1980five,schwarz1985mass,gooding1997care,capinera2012sweetpotato,zheng2015standard,carvalho2014mass}. These differences present two obstacles to SIT that will be testable within the scope of my model. First, because of the long life cycle, \textit{I. scapularis} SIT might simply take too long. Second, because sterile ticks won't disperse much on their own, manually blanketing the sterile ticks across every part of the control area may be necessary \cite{lance2021mass,alphey2010sterile}. To assess these concerns, I use the model first to estimate program durations on a single control patch over a variety of parameter values, and then expand the model to a metapopulation and test the relative performances of poorly and thoroughly distributed sterile release layouts.

\subsection{\textit{I. scapularis} life course}

\begin{figure}[t]
\centering
\includegraphics[scale=0.75]{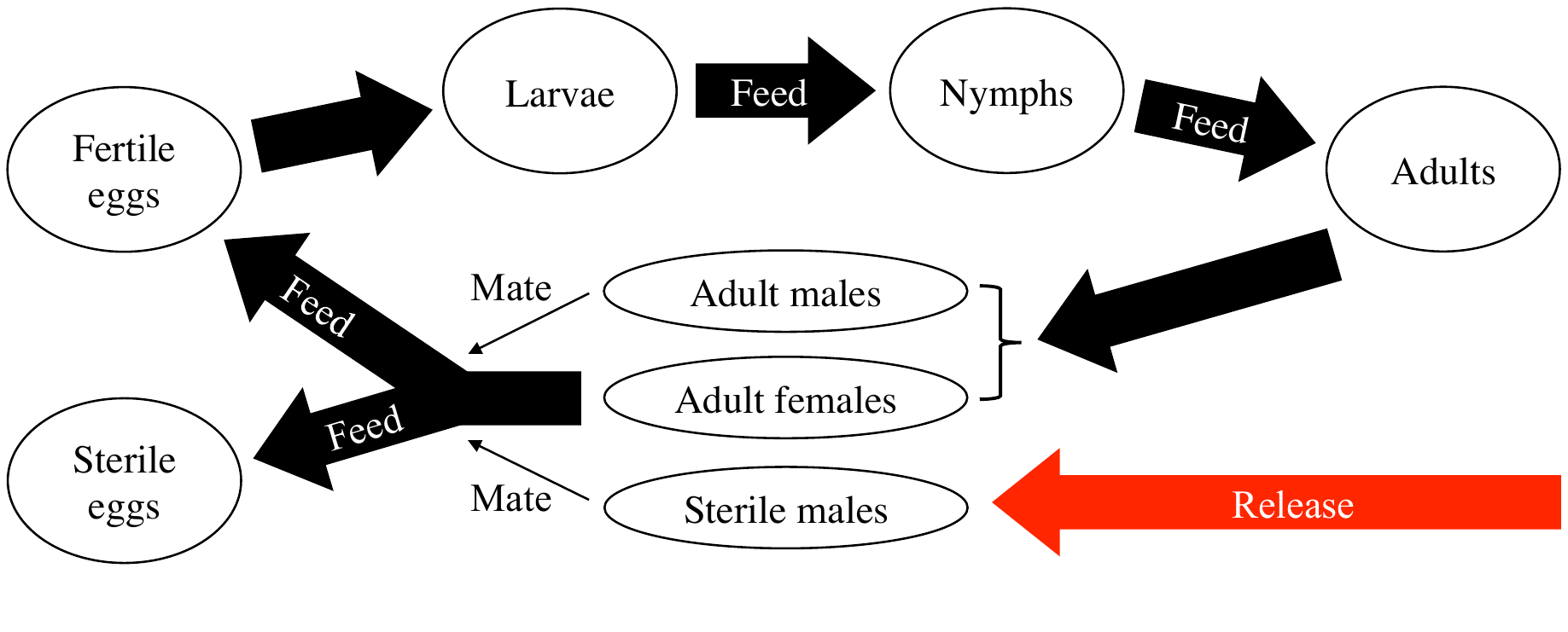}
\caption{The life course of \textit{I. scapularis}, which typically takes two years (sometimes up to four) in the wild. Larvae and nymphs drop off their hosts and molt after feeding. Female ticks also drop off their hosts to lay eggs, after which they die. Sterile male release causes a fraction of the total eggs to be sterile.
}
\label{course}
\end{figure}

The life course of wild \textit{I. scapularis} typically takes two years (though some ticks may survive to complete it in three or four) and is divided into four stages: egg, larva, nymph, and adult (Figure \ref{course}) \cite{yuval1990duration,halsey2018spatial}. Larvae and nymphs must find hosts and feed to molt to the next stages. Adult females must find a third host and feed during or after mating, which can occur on or off host \cite{fish1993population,troughton2007life}. Females attempt to mate and feed in the fall, and if successful will overwinter before laying eggs; otherwise they have a window in the spring to mate (if needed) and find their last host \cite{sandberg1992comprehensive,kocan2015insights,daniels1989seasonal}. Total feeding time is no more than 23 days, and the ticks do not remain on hosts between feedings \cite{kocan2015insights,randolph2013ecology}. Females and most males mate with only one other tick \cite{goltz2012ecology,troughton2007life,kocan2015insights}.

\section{Methods}

\subsection{Geometric growth model with hyperbolic density-dependence}

Density-dependent population growth has been observed in \textit{I. scapularis}, due at least in part to carrying capacities of tick hosts \cite{gaff2020lymesim,fish1993population} (Supplementary materials of \cite{gaff2020lymesim}). As the core of my deterministic, discrete-time population model, I follow Prout (1978) in using hyperbolic density-dependent growth, which is equivalent to a logistic model measured at integer values and generally conforms well to observations \cite{prout1978joint,poulsen1979model,randolph2013ecology}. However, the model does omit density-dependence in fertility; I will add this later when I introduce Allee effects.

The hyperbolic density-dependence model yields a simple S-curve for population growth, with the highest growth rate at moderate population density. Growth of a high-density population flattens out as it approaches a value known as the ``carrying capacity,'' which I fix at $1$ \cite{prout1978joint}. The model is given by:
\begin{equation}\label{hdd}
N_{t+1} = \frac{\lambda N_t}{1+(\lambda-1)N_t},
\end{equation}
where $N_t$ is the population of generation $t$ (as a proportion of the carrying capacity) and $\lambda$ is the density-independent population growth rate. In particular, $N_t$ represents the ticks of the $t$-th generation that survive to their mating season, when sterile release occurs.

Notice that the model includes only one population of ticks, whereas two main cohorts exist in the wild because the typical life cycle of \textit{I. scapularis} is two years. Since the model is independent of absolute population size, this should make no difference.

\begin{table}[t]
\begin{center}
\caption{Parameters \& Variables}
\begin{tabular}{c|l|c|c|}
 & Meaning & Value range & Meta. value \\
\toprule
$N_t$ & wild pop. of $t$-th generation at time of release & $N_0\in[0.1,0.7]$ & $0.4$ \\
$\lambda$ & density-independent growth constant & $[1.15,1.77]$ & $1.33$ \\
$t$ & time (unit varies: generations or years) & & \\
\midrule
$S$ & sterile male release rate (per year) & & \\
$W_N$ & wild proportion of adult males after release & & \\
$b$ & average eggs laid per female & & \\
$s$ & density-independent survival rate & & \\
\midrule
$T$ & Allee threshold & $[0.01,0.2]$ & $0.1$ \\
$A$ & strength of mate-finding Allee effect & $[0.00075,0.077]$ & $0.0665$ \\
$X_N$ & proportion of female ticks that find a mate & & \\
\midrule
$\mu$ & migration rate & $[0,\text{max}]$ & $0.25$ \\
\bottomrule
\end{tabular}\label{table}
\end{center}
\end{table}

\subsection{Sterile male release}

The first and still standard approach to modeling SIT, proposed by Knipling (1955)  \cite{knipling1955possibilities}, is to scale mating successes by the fraction of males that are wild (i.e. fertile), which I denote $W_N$: 
\begin{equation}\label{W}
W_N = \frac{\frac{1}{2}N_t}{\frac{1}{2}N_t+S},
\end{equation}
where $\frac{1}{2}N_t$ is the wild males, because ticks have an equal sex ratio \cite{sandberg1992comprehensive}, and $S$ is the number of sterile males released per mating season (i.e. per year) as a proportion of the wild population carrying capacity. In order to incorporate this into the model, first observe the natural relation:
\begin{subequations}
\begin{align}
\lambda &= \frac{1}{2}bs \\[4pt]
\Rightarrow\;\; \lambda &\propto b, \label{b}
\end{align}
\end{subequations}
where $b$ is the average number of eggs laid per female (which implicitly includes mate-finding success rate), $\frac{1}{2}$ is the proportion of hatches that are female, and $s$ is the density-independent survival rate from egg to life course completion \cite{prout1978joint,sandberg1992comprehensive}. Knipling's approach then instructs that $b$, hence $\lambda$, be scaled by $W_N$. One might also consider altering the carrying capacity, following Prout (1978) \cite{prout1978joint}; however, known sources of density-dependent mortality in \textit{I. scapularis} are not likely to be influenced by sterile release \cite{vail1997density,gaff2020lymesim,samish1999pathogens} (Supplementary materials of \cite{gaff2020lymesim}), so I do not alter this here. Thus incorporating $W_N$ into Equation \ref{hdd} simply yields:
\begin{equation}\label{release}
N_{t+1} = N_t \cdot \frac{\lambda W_N}{1+(\lambda W_N-1)N_t}.
\end{equation}

The model assumes sterile ticks are released en masse (at the beginning of the mating season). This is consistent with defining $N_t$ as the wild population at the time of release, as stated in Table \ref{table}.

\subsection{Mate-finding Allee effect}\label{section:allee}

Hyperbolic density-dependence handles the dynamics of moderate and large populations well, but not small populations. If a population is small, we might find a reduced per-capita growth rate in the wild (for instance, because of mate-finding failure); this is captured by Allee effects \cite{taylor2005allee}. A ``strong'' mate-finding Allee effect is to be expected in ixodid ticks \cite{kada2017impact,tobin2007invasion}, although it has not been demonstrated and likely has a low ``Allee threshold.'' A strong Allee effect simply means that there exists some population density (the Allee threshold) below which decline toward extinction occurs naturally (in this case by mate-finding failure) \cite{taylor2005allee}. I could set this threshold at two ticks, as has been done previously \cite{weidhaas1983basic}; but it is likely higher, although on-host mating (and perhaps other factors) may keep it fairly low \cite{nadolny2016hitchhiker}. An advantage of including an Allee threshold in the model is that it provides a practical target population for an SIT program \cite{liebhold2003allee,boukal2009modelling}, while exerting minimal influence on population dynamics during control (see Section \ref{section:analysis}). A further advantage of the Allee threshold is that it will allow the addition of immigration without eliminating the possibility of stable near-extinction. With immigration but no Allee threshold, the only equilibrium is at the carrying capacity.

Kada \textit{et al.}\;(2017) introduce an inverse density-dependent term to fertility representing the proportion of females that are able to find a mate: $\frac{1}{2}N_t / (\frac{1}{2}N_t+A)$, where $\frac{1}{2}N_t$ is the number of males and $A$ indicates the strength of the Allee effect \cite{kada2017impact}. I must modify this term to account for the introduction of sterile males, which facilitate mate-finding; this is easily accomplished by adding $S$ to each $\frac{1}{2}N_t$. Then I define:
\begin{equation}\label{X}
X_N = \frac{\frac{1}{2}N_t+S}{\frac{1}{2}N_t+S+A},
\end{equation}
the proportion of female ticks that are able to find a mate. Notice that the smaller the population, the greater the influence $A$ exerts, as desired.

Finally, this factor must be applied to $b$, since $b$ implicitly included mate-finding as a constant element. As in the last section, this is accomplished by multiplying each $\lambda$ by $X_N$. Also as above, I choose not to allow the carrying capacity to vary from 1, since a mate-finding Allee effect only significantly impacts small populations. Then incorporating $X_N$ into Equation \ref{release} yields:
\begin{equation}\label{Allee}
N_{t+1} = N_t \cdot \frac{\lambda W_N X_N}{1+(\lambda W_N X_N-1)N_t}.
\end{equation}

Later I will use Allee threshold, which I denote $T$, instead of Allee strength ($A$) for convenience. Allee strength can easily be computed from Allee threshold by the relation (Appendix \ref{appendix:a3}):
\begin{equation}\label{A}
A = \frac{1}{2}(\lambda-1)T.
\end{equation}
Note that the Allee threshold is only the true threshold of population decline in the absence of both sterile release and migration.

\subsection{Analysis of no-migration model}\label{section:analysis}

Substituting for $W_N$ and $X_N$ in Equation \ref{Allee} using Equations \ref{W} and \ref{X} leads to some canceling (Appendix \ref{appendix:a1}), yielding the final no-migration model:
\begin{equation}\label{final}
N_{t+1} = N_t \cdot \frac{1}{N_t+\frac{1-N_t}{\frac{1}{2}\lambda N_t}\left(\frac{1}{2}N_t+A+S\right)}.
\end{equation}

Notice that sterile release ($S$) creates the equivalent of an increase in the strength of the Allee effect ($A$); this is expected, as sterile release acts like an Allee effect \cite{boukal2009modelling}. Sterile release values will typically be far higher than Allee strength values, so the Allee effect will have minimal impact during control (as claimed in Section \ref{section:allee}).
Solving Equation \ref{final} for equilibria (Appendix \ref{appendix:a2}), denoted $N^*$, yields $N^*=0$, $N^*=1$, or:\\
\begin{equation}\label{equil}
N^* = \frac{A+S}{\frac{1}{2}(\lambda-1)}.
\end{equation}

\begin{figure}[t]
\centering
\includegraphics[scale=0.6]{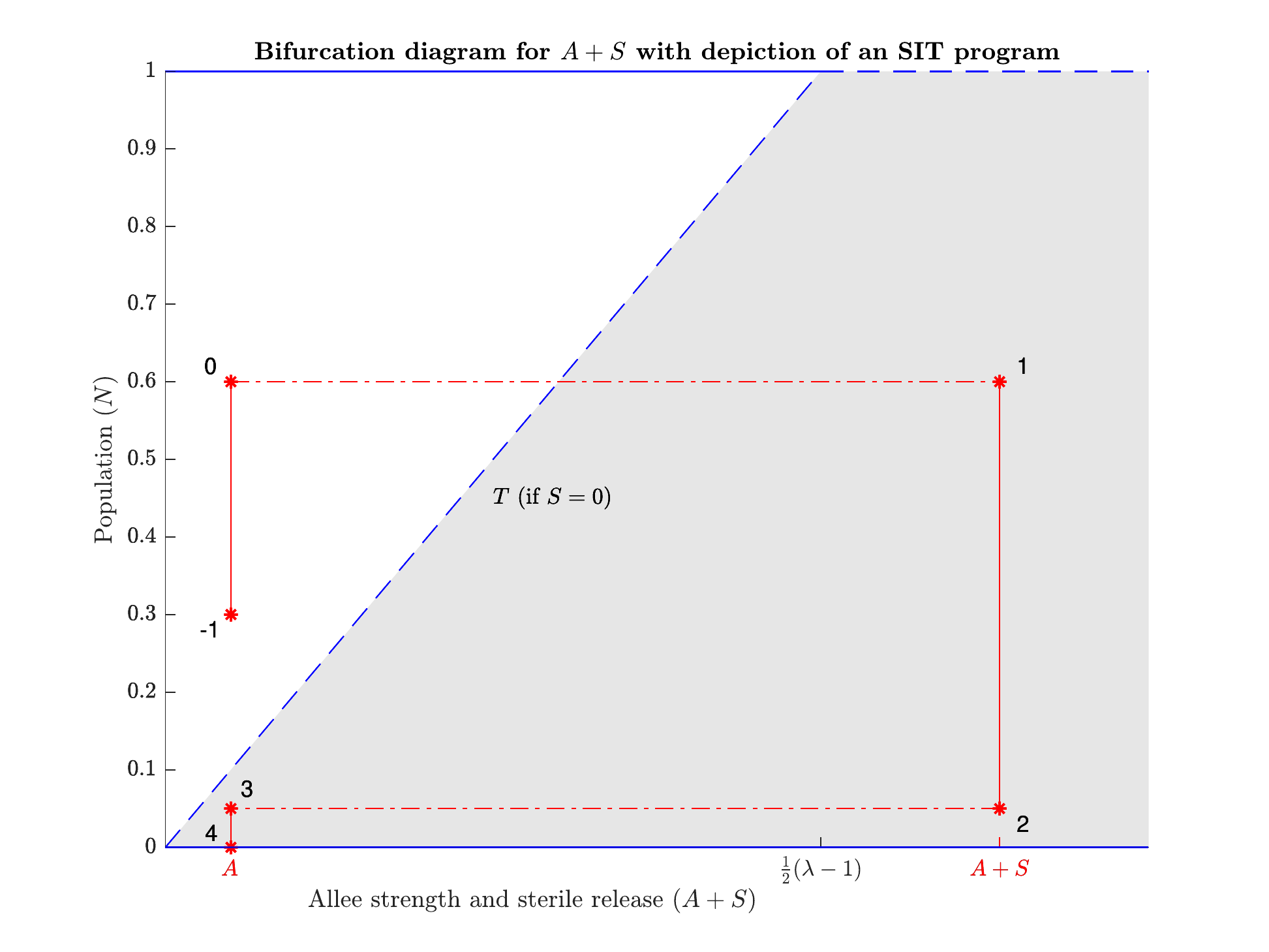}
\caption{Bifurcation diagram for Equation \ref{final}; solid blue lines indicate stable equilibria and dashed blue lines indicate unstable equilibria. Population decline toward extinction occurs in the gray region. In the absence of sterile release, the angled line is the Allee threshold ($T$). Red lines depict the course of a typical SIT program, with sterile release commencing at point $0$ and terminating at point $2$.}
\label{bifurc}
\end{figure}

These thresholds form a transcritical bifurcation (Figure \ref{bifurc}). Extinction ($N=0$) and the carrying capacity ($N=1$) are generally stable equilibria (indicated by solid blue lines), while the threshold of population decline is an unstable equilibrium (indicated by a dashed blue line). In the absence of sterile release, this last is the Allee threshold. Gray fill shows the region of population decline. Notice that the carrying capacity equilibrium at $1$ persists to the right as an unstable equilibrium, which is not biologically realistic (population decline should occur); this issue derives from using hyperbolic density-dependence with $\lambda$ effectively less than $1$. My model can't be trusted for population values near or above this carrying capacity equilibrium. In particular, sterile release requirements predicted by the model begin to climb noticeably above $N=0.7$, as a bogus asymptote at $N=1$ is approached.

I can now more easily describe the process of an SIT program: some release $S>0$ is applied to shift the population into the area of decline (gray in Figure \ref{bifurc}), until the population is low enough that it will remain in the gray area without release (i.e. $S=0$). An example of this approach is depicted by the red lines in Figure \ref{bifurc}. The population is initially increasing (from point $-1$ toward point $0$) before control begins; the value of $A+S$ is simply $A$, as indicated in red. At point $0$, implementation of release begins and increases $A+S$ dramatically to a value in the gray region (dash-dot red line to point $1$). The population now declines. Once the population is sufficiently reduced (point $2$), release is halted and $A+S$ returns to the value of $A$ (dash-dot line to point $3$). Notice that the population is still in the gray area even in the absence of sterile release. The Allee effect therefore eradicates the remnants of the population without release (points $3$ to $4$).

\subsection{Simple migration}\label{section:mig}

The flow of ticks into and out of a control area could hinder SIT (although ticks' low migration reduces this) \cite{lance2021mass}. Thus, I wish to include migration in the model before estimating program durations. In this simple migration implementation, the control area is modeled as a single large patch.

Tick migration is unusual in that it is almost entirely host-driven \cite{heylen2010contrasting,nadolny2018modelling}. I model this by simply having a fixed proportion of ticks migrate each generation. Thus I have density-dependent emigration from the control area, as discussed by Prout (1978) \cite{prout1978joint}. However, inspired by Kada \textit{et al.}\;(2017), I assume that there is a source patch at carrying capacity equilibrium \cite{kada2017impact}, so immigration into the control area is constant. I will modify this later to model metapopulation migration.

In the present model, sterile release is the first event each generation, and its value $S$ is integrated into subsequent density-dependent population growth. Ideally my model would also integrate migration and density-dependent mortality, since migration occurs multiple times over the tick life course; however, for simplicity I separate these factors. Migration rates will be low enough as to minimally affect density-dependent mortality most of the time, so this is a reasonable approximation. An exception is when the controlled population has become so small that immigration significantly increases it over the course of a single generation; this circumstance is transient, but it could cause problems if the initial population is too small, as noted in Section \ref{section:N0}.

Splicing migration in between release and population growth is the next best option, because ticks' final host (which immediately follows release) is typically the most mobile \cite{halsey2020maintenance}, but for simplicity I place migration last. This simplification will impact the first and last generations of a program and slightly reduce overall migration; I will allow migration to range higher to approximately correct for this (see Section \ref{section:mu}). Equation \ref{final} becomes:\\
\begin{equation}\label{mig}
N_{t+1} = N_t \cdot \frac{1-\mu}{N_t+\frac{1-N_t}{\frac{1}{2}\lambda N_t}\left(\frac{1}{2}N_t+A+S\right)} + \mu,
\end{equation}\\
where $\mu$ is the migration rate (per generation) between patches. I have assumed immigration and emigration cancel each other out exactly when the controlled population is at its carrying capacity, forcing the value of equilibrium of the source patch to equal that of the control patch, which is $1$.

Migration complicates the lower two equilibria in Figure \ref{bifurc} a bit. If immigration isn't too high, the threshold of population decline is simply a bit below the Allee threshold, and the lowest stable equilibrium is a little higher than complete extinction. However, if immigration is sufficiently high, it will overwhelm the Allee threshold, leaving carrying capacity as the only equilibrium \cite{drake2006allee}. This is discussed further when I set the range of values for $\mu$ in Section \ref{section:mu}.

\subsection{Migration generalized to a metapopulation}\label{section:meta}

With a metapopulation model, the control area is described by many small patches instead of one large patch; each small patch is modeled individually like the single patch was modeled before.

I translate Equation \ref{mig} to a metapopulation model by modifying the immigration term; emigration need not change. Recall that immigration previously came from a stable source patch. I now take immigration to be a sum of emigrations from nearby patches. In particular, I use a grid of patches and only single-step, non-diagonal migration, so immigration to a given patch is one-fourth the sum of emigrations from the four adjacent patches. For each patch this can be written:
\begin{equation}\label{meta-imm}
N_{t+1} = N_t \cdot \frac{1-\mu}{N_t+\frac{1-N_t}{\frac{1}{2}\lambda N_t}\left(\frac{1}{2}N_t+A+S\right)} + \frac{1}{4}\sum_{i=1}^4 \text{adj}_i\cdot\mu,
\end{equation}

where the $\text{adj}_i$'s represent the four patches adjacent to the given patch. For individual patches this model retains qualitatively the same equilibria as the simple migration model.

\subsection{Parameter values}\label{section:ranges}

This section provides references and rationale for the parameter values and ranges used and given in Table \ref{table}.

\subsubsection{\texorpdfstring{$N_0$}{N0}}\label{section:N0}

I consider a minimum initial population size of $N_0=0.1$, for two main reasons. First, a value smaller than this could simply be emigration overflowing from a nearby population but not establishing. Second, as discussed in Section \ref{section:mig}, the model ought to have density-dependent mortality be temporally interwoven with migration for small populations, which the model does not. I cap populations at $N_0=0.7$, also for two reasons. First, \textit{I. scapularis} SIT is unlikely to be attempted on fully established, previously uncontrolled populations (at least initially): a standard approach is to knock down the population with other controls and then apply SIT (a form of AW-IPM) \cite{klassen2021history,knipling1955possibilities,stafford2017integrated}. Second, with my model sterile release requirements grow asymptotically as a bogus equilibrium at $N_t=1$ is approached, as discussed in Section \ref{section:analysis}. A central value of $0.4$ is used with the metapopulation model.

\subsubsection{\texorpdfstring{$\lambda$}{lambda}}

Awerbuch-Friedlander \textit{et al.}\;(2004) compute parameters similar to $\lambda$ based on fieldwork by Sandberg \textit{et al.}\;(1992); their results equate to $\lambda\approx1.33$ at population sizes minimally influenced by the Allee effect \cite{awerbuch2005role,sandberg1992comprehensive}. In examining program durations I test the broader range $\lambda\in[1.33^{1/2},\,1.33^2]\approx[1.15,\,1.77]$. Extending this range higher or lower does not significantly impact the results.

\subsubsection{\texorpdfstring{$T$}{T}}\label{section:T}

Average Allee threshold values for ticks are not known. Tobin \textit{et al.}\;(2007) use $0.2$ as a general species-independent Allee threshold value \cite{tobin2007invasion}; it is expected that the Allee threshold is rather low for ixodid ticks \cite{nadolny2016hitchhiker}, so in examining program durations I test the range $T\in[0.01,\,0.2]$. Allee thresholds can be quite variable across both region and time \cite{tobin2007invasion}, so it is reasonable to use such a large range. Note that in estimating program durations the program goal will be to reduce the population into a state of natural decline, so I cannot test $T=0$ as this eliminates the possibility of success. For the metapopulation tests I choose the moderately high value of $T=0.1$; this choice has negligible effect (see Section \ref{section:analysis}), and is made simply to allow the use of moderately high migration. Using $T$, $\lambda$, and Equation \ref{A}, one can also determine the values for $A$ (given in Table \ref{table}).

\subsubsection{\texorpdfstring{$\mu$}{mu}}\label{section:mu}

Migration rate, $\mu$, should be quite low, since ticks are nearly immobile off-host and spend well under a month on hosts during their two or more year life course, and often only a week or so on a significantly mobile host \cite{falco1991horizontal,carroll1996dispersal,kocan2015insights,gaff2020lymesim,halsey2020maintenance} (Supplementary materials of \cite{gaff2020lymesim}). What constitutes low migration depends on the spatial scale, since smaller patches or control regions experience higher migration rates \cite{alphey2010sterile}.

As noted in Section \ref{section:mig}, if $\mu$ is too high then sterile release will be useless in achieving stable near-extinction (or extinction). While such significant migration may be required for range expansion, at least as a stochastic event, in my estimation of ideal program durations I must assume program completion is possible (i.e. stable extinction or near-extinction), so I take migration to be sufficiently low. Then the maximum value for $\mu$ must certainly be less than $T$, else immigration from the source patch would meet the Allee threshold every generation. The exact upper limit formula is extremely complicated, and I need not explicitly find it. Suffice it to say that $\mu$ must be quite small, with maximum values on the order of $\frac{1}{10}T$. This is not a problem, given the earlier observations on the low migration of \textit{I. scapularis}. I take $\mu$ values right up to the limit to mitigate the migration reduction from the imperfect order of events introduced and discussed in Section \ref{section:mig}: the limit is naturally higher to match the order used.

In the metapopulation framework $\mu$ ought to be much larger, given the fine patch scale (see Section \ref{section:mu}). The higher $\mu$, the smaller is the implied patch size, allowing a finer-grained and more plausible layout for the less thorough release distribution. Presumably I must take $\mu$ to be below $0.8$, though: this is the value at which ticks disperse evenly between a given patch and the four surrounding patches. Indeed, I expect patches to be large enough that $\mu$ is well below $0.8$, else my simplification that migration occurs only between adjacent patches would be questionable. This gives a sense of the $\mu$ value to expect, but for consistency we determine the value roughly as we did in the program duration tests. To do this, we compute the total emigration rate over the border of a uniform metapopulation control area as a function of $\mu$; given $T=0.1$ from the previous section, the maximum allowed emigration is around $0.01$, yielding approximately $\mu=0.25$. We use this moderately high value to give the less than thorough release layout a fair chance.
\pagebreak 
\section{Results}

\subsection{Ideal program durations}\label{section:times}
\vspace{-.5\baselineskip} 
\subsubsection{Preliminary examples}
\vspace{-.5\baselineskip} 
To study ideal program durations, I use my model to simulate programs for a variety of parameter values. It suffices to use simple migration as given by Equation \ref{mig} and to vary parameters within the ranges given in Table \ref{table}, with the exception of release rate, which is given unlimited range. For simplicity, though, I assume release rate remains constant over the course of each program.

\begin{figure}[t]
\centering
\includegraphics[scale=0.7]{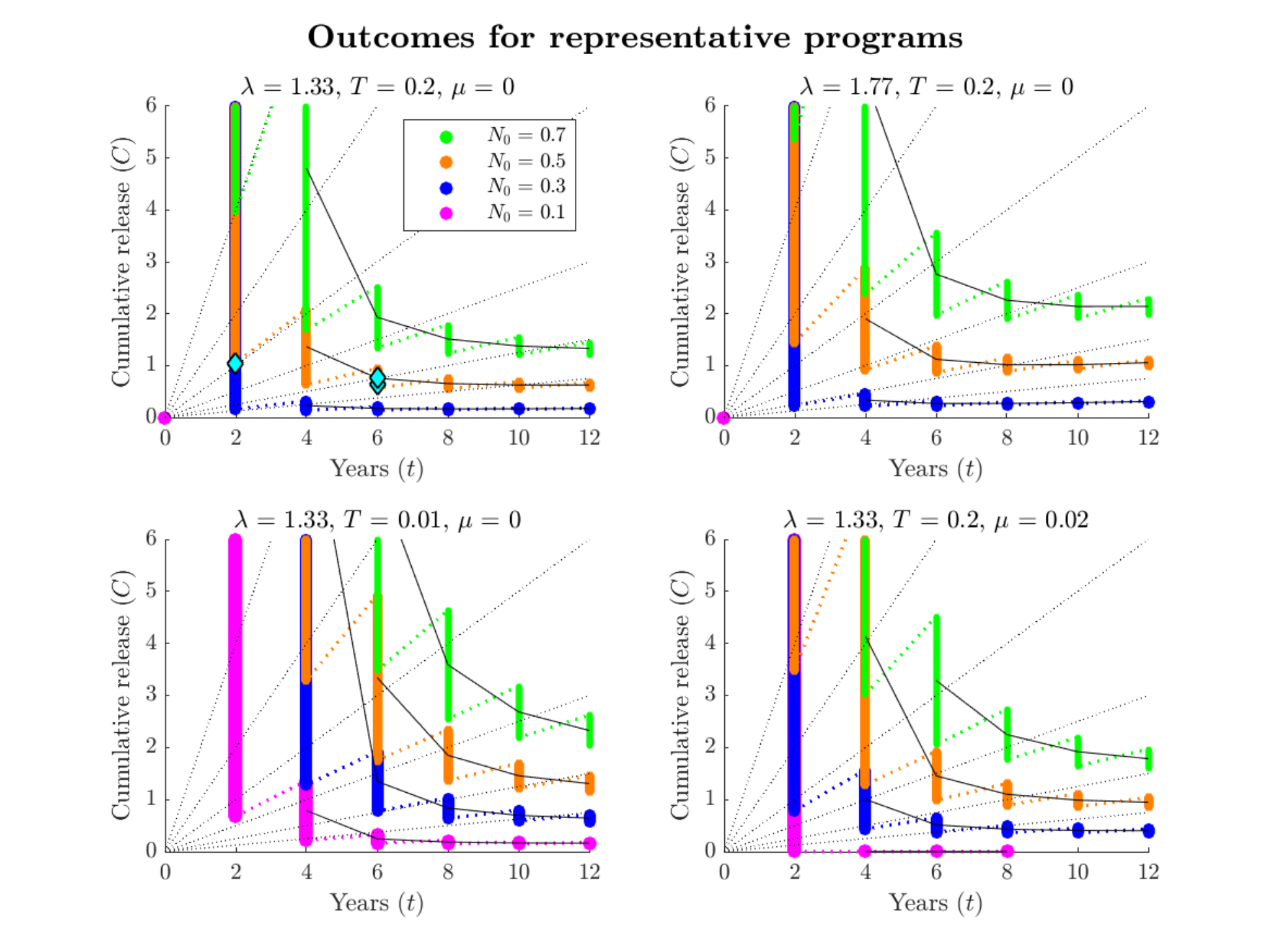}
\caption{SIT programs are simulated for varying release rate (with other parameters fixed) and each outcome is plotted as a circle; these circles overlap to form ``stacks'' of outcomes. Colors indicate different initial populations ($N_0$) and each subplot uses a different set of values for the other parameters. Colored dotted lines indicate jumps in outcome duration as release rate ($S=C/t$) varies. A few arbitrary release rates are given as black dotted lines to help convey the nature of the figure. Black lines traverse midpoints of stacks. Examples of ideal programs (by different criteria) are shown as cyan diamonds in the upper-left subplot.}
\label{stacks}
\end{figure}

Simulation outcomes for some representative parameter values are plotted as circles (which overlap to form thick lines) in Figure \ref{stacks}, along with the parameter values used. The variables on the axes are cumulative sterile release compounded over each program (in units of wild population carrying capacity) and time to program completion (now given in years). Both of these are dependent on release rate: each simulation yields the time and cumulative release required to bringing the population into natural decline toward stable extinction or near-extinction. Cumulative release is a continuous variable and time required is discrete, so multiple releases correspond to each outcome duration; with release rate varying continuously, the result is the outcome ``stacks'' in Figure \ref{stacks}.

It may help elucidate Figure \ref{stacks} to consider some example simulations: say, the green stacks in the upper-left subplot. All parameters except release rate are fixed; different parameter values are given as different colors or in different subplots. As shown, the values corresponding to the green stacks of the upper-left subplot are $N_0=0.7$, $\lambda=1.33$, $T=0.2$, and $\mu=0$. For a given release rate, the model can be used to simulate a program to completion; this will take some number of years and some cumulative release $C=S\cdot t$. The result is plotted on the subplot as a green dot, perhaps at $(4,4)$. Varying release rate a little and simulating again changes the outcome and moves the dot up or down a bit on the stack. If release rate is varied substantially, the outcome may reach the top or bottom of the stack and jump to the next along a green dotted line; in other words, the duration of the program changes. Depicting outcomes from all relevant release rates produces the green stacks shown.

Notice that the variables on the axes ($C$ and $t$) make release rate the slope from the origin to an outcome ($S=C/t$), so increasing release rate smoothly moves through the stacks from right to left and bottom to top. A few arbitrary release rates are shown as black dotted lines to help convey this.

\subsubsection{Ideal release rates}\label{section:criteria}

Depending on priorities, an ideal program might minimize time required ($t$), cumulative release ($C$), or both. I assume that the last of these is most likely, as minimizing only $t$ often requires dramatically increased release and minimizing only $C$ often means significantly increasing program duration for minimal improvement (see Figure \ref{stacks} for examples). Minimizing both $C$ and $t$ can be done in any number of ways. I select perhaps the neatest: minimization of the product $C\cdot t$. An example of an ideal outcome by this criterion is given in the upper-left subplot of Figure \ref{stacks} for $N_0=0.5$ (orange stacks); it appears as the leftmost cyan diamond.

It may be unrealistic to target the very bottom of a stack, as minimizing $C\cdot t$ requires. Any error or variation down on the stack would jump the outcome to the top of the next stack to the right, which would be considerably less desirable (Figure \ref{stacks}). Attempting to monitor the wild population with sufficient accuracy may be a possibility, but we assume here that that is not practicable \cite{mooring1995efficiency,daniels1989seasonal}. Thus, simply minimizing $C\cdot t$ is a risky criterion to use.

A natural approach to ensuring program completion might be to continue release for an extra two-year generation beyond what the model requires. This equates to minimizing the product $C\cdot(t+2)$; for $N_0=0.5$ in the upper-left subplot of Figure \ref{stacks}, this yields the lower-right cyan diamond shown (largely obscured).

Another possible method for ensuring completion might be to consider only stack midpoints, given by black lines in Figure \ref{stacks} (midlines do not extend to the leftmost stacks because these stacks are infinitely tall). Minimizing $C\cdot t$ along the midline for $N_0=0.5$ in the upper-left subplot of Figure \ref{stacks} yields the upper-right cyan diamond shown.

All three of these criteria are robust to linear scaling of release, since they simply minimize a product including the quantity scaled. The importance of this is discussed later (Section \ref{section:secondary}).

\subsubsection{Aggregated duration estimation}

\begin{figure}[t]
\centering
\includegraphics[scale=0.7]{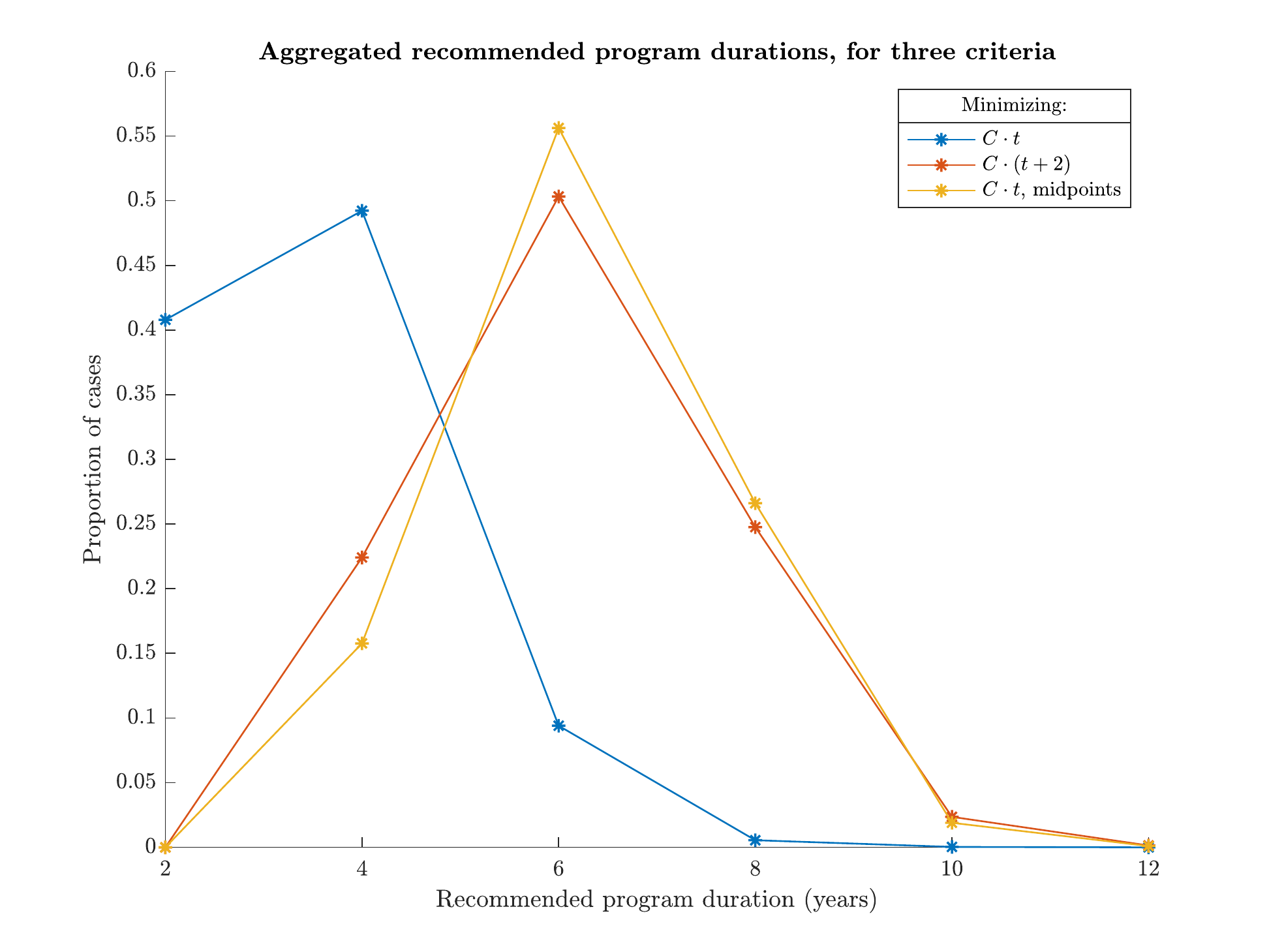}
\caption{SIT programs are simulated over a Latin Hypercube Sample of parameter values and resulting program durations are graphed in aggregate. One plot line is given for each of three different ideal release rate criteria used. The LHS is generated with 20,000-bins over the parameter ranges $\lambda\in[1.15,1.77]$, $T\in[0.01,0.2]$, $N_0\in[0.1,0.7]$, $\mu\in[0,\text{max}]$. The legend shows the three terms minimized to determine ideal release rates (see Section \ref{section:criteria}). Cases that are uncontrollable or where no release is needed are omitted from proportion calculation.}
\label{times}
\end{figure}

To analyze recommended program durations using the three criteria above, I generate a 20,000-bin Latin Hypercube Sample of parameter values over the ranges found in Section \ref{section:ranges} (given in Table \ref{table}). Aggregated results are given in Figure \ref{times}, with one line per ideal release rate criterion. Cases that are uncontrollable (53.6\% of total) or where no release is needed (1.3\% of total) are omitted from proportion calculation. The latter result from immigration exceeding the upper limit (see Section \ref{section:mu}), which I cut out at this stage rather than attempting to avoid earlier.

Figure \ref{times} shows that, for all three criteria, well under $5\%$ of cases require more than eight years. As expected from Figure \ref{stacks}, simply minimizing $C\cdot t$ yields the shortest programs of the three; but this criterion is risky, as discussed in Section \ref{section:criteria}. The two safer approaches yield nearly identical recommendations, with just over half of the cases coming in at six years and around one-fourth at eight years.

\subsubsection{Secondary observations}\label{section:secondary}

It is common for models of SIT to considerably under-predict release requirements, and indeed my model yields values well below the expected range: a standard release rate is 10 times the wild population or more per mating season \cite{benedict2021sterile}, whereas my mean recommended release rates are less than $0.75$ for all ideal release criteria. The primary factor in this under-prediction is most likely reduced competitive ability of the sterile males resulting from the sterilization and release process \cite{benedict2021sterile,merino2011targeting,whitten2021misconceptions,benedict2003first}. Not knowing the magnitude of this effect in \textit{I. scapularis}, all I can observe is that reduced competitive ability should approximately linearly scale the number of ticks needed \cite{barclay2021mathematical}. It would therefore be unreasonable to estimate release as I estimate duration. I only take my results for the latter to be plausible because my ideal program criteria are robust to linear scaling of release (see Section \ref{section:criteria}).

I also do not present the individual effects of each parameter on program duration, for three reasons. First, parameter values are unlikely to be known in a given control area. Second, parameters will likely vary within any sizeable control region \cite{tran2021spatio,gaff2020lymesim}. Third, studying individual parameter effects would require a substantially more elaborate analysis of the model. However, interested readers can glean some relations from Figure \ref{stacks}. For instance, increasing $N_0$ appears to move the stacks roughly linearly away from the origin.

\subsection{Poorly vs. thoroughly distributed sterile release}\label{section:meta-result}

\subsubsection{Scenarios}

A prominent concern with \textit{I. scapularis} SIT is the released ticks' inability to disperse on their own \cite{lance2021mass,knipling1955possibilities,falco1991horizontal,carroll1996dispersal}. I therefore use my model to test the effectiveness of release that is not distributed to all parts of the control area, in contrast to thorough distribution, presuming that the former would be substantially easier and cheaper, and thus preferable unless the two yield dramatically different results \cite{dowell2021supply}. To simulate localized control, I take the control area to be part of a larger uniformly tick-populated region; low tick migration and the Allee effect bode well for the persistence of such local eradication \cite{lance2021mass,yamanaka2009spatially}.

\begin{figure}[t]
\centering
\begin{subfigure}{0.45\textwidth}
\centering
\includegraphics[scale=0.4]{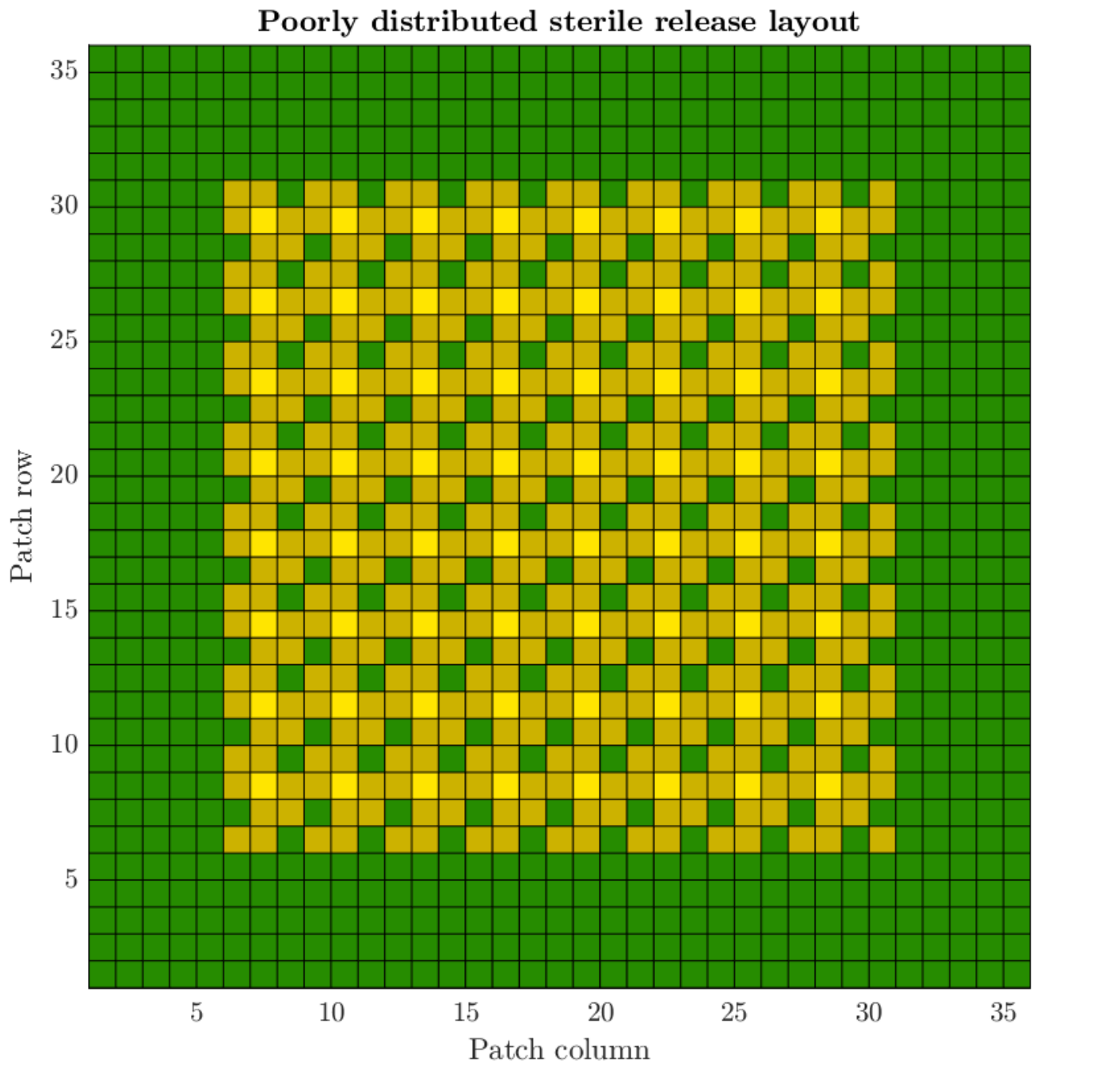}
\caption{}
\label{layout-poor}
\end{subfigure}
\begin{subfigure}{0.45\textwidth}
\centering
\includegraphics[scale=0.4]{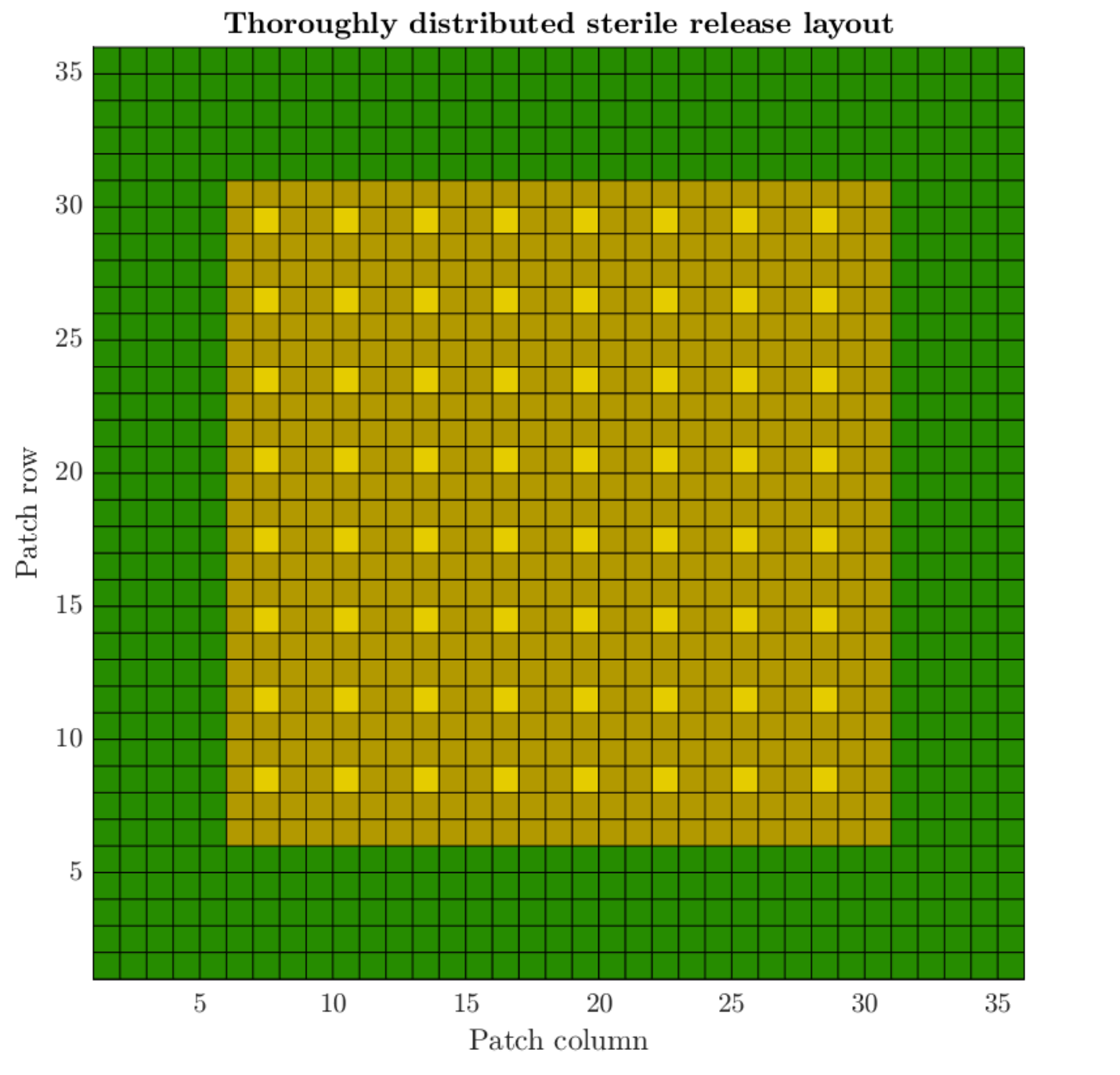}
\caption{}
\label{layout-thor}
\end{subfigure}
\begin{subfigure}{0.45\textwidth}
\centering
\includegraphics[scale=0.4]{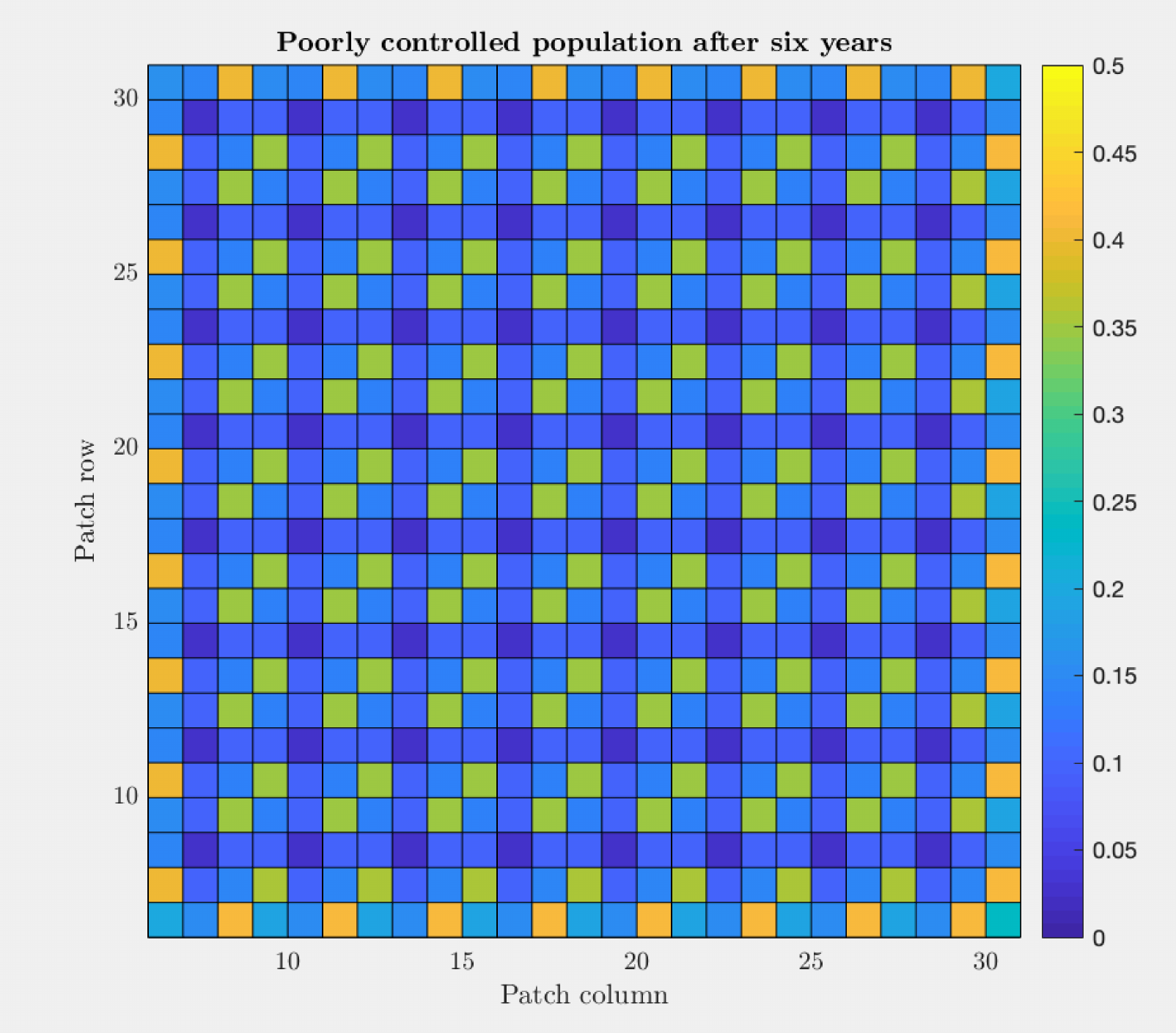}
\caption{}
\label{poor-result}
\end{subfigure}
\begin{subfigure}{0.45\textwidth}
\centering
\includegraphics[scale=0.4]{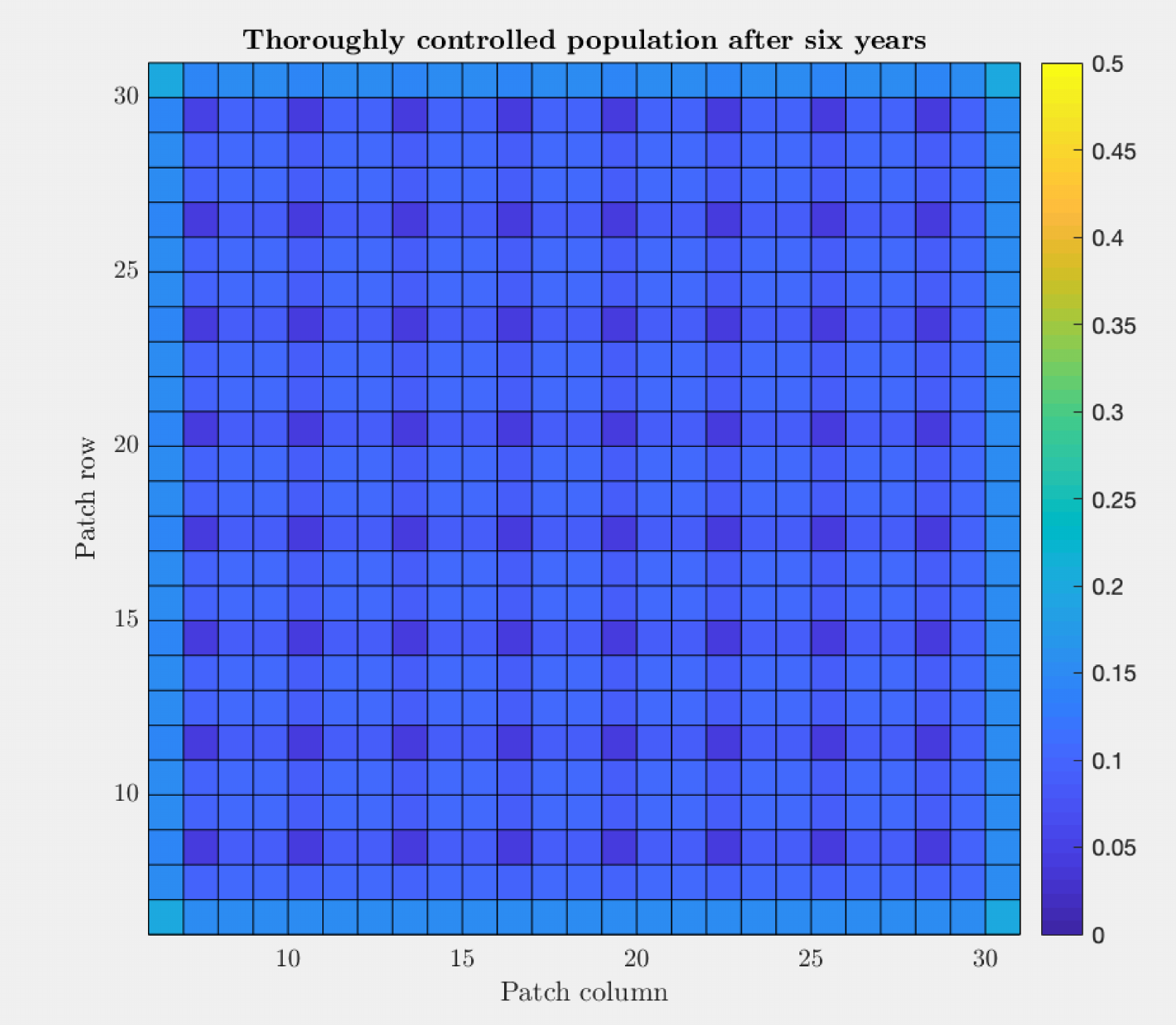}
\caption{}
\label{thor-result}
\end{subfigure}
\caption{Poorly and thoroughly distributed sterile release layouts and their effects after six years are shown. Total release is the same in both scenarios. Parameter values used are $N_0=0.4$, $\lambda=1.33$, $T=0.1$, $\mu=0.25$, and mean $S=0.2$. \textbf{(A)} The ``poorly'' distributed release layout, with some of the surrounding tick-populated region shown for reference. Release occurs only on the yellow patches, with brighter yellow indicating double-release patches. \textbf{(B)} The ``thoroughly'' distributed release layout, with some of the surrounding tick-populated region shown for reference. Sterile ticks are released on all yellow patches, with bright yellow indicating double-release patches. \textbf{(C)} The wild population in the control area after six years of poorly distributed release. \textbf{(D)} The wild population in the control area after six years of thoroughly distributed release.}
\end{figure}

I test two release layouts, both using the same total sterile tick release rate. These are shown in Figures \ref{layout-poor} and \ref{layout-thor}. Release occurs on yellow patches, with brighter yellow indicating higher release. Notice that even the ``poor'' distribution is still rather thorough, providing a plausible comparison scenario. The poor layout used is intended as a representative case of non-aerial release, with $2\times2$ regions of release distributed throughout \cite{dowell2021supply}. For simplicity, the thorough distribution is uniform, except for some brighter yellow double-release patches matching those of the poor distribution.

To test effectiveness of sterile release for the two layouts, I simulate each for six years using the central parameter values selected in Section \ref{section:ranges} (given in Table \ref{table}). The rates of release recommended by the safer criteria in Section \ref{section:times} are just on either side of $S=0.2$ (computing $\mu$ as in Section \ref{section:mu}), so I use $S=0.2$ as the mean rate for both control areas. Note that $S=0.2$ is low compared to what one would expect to need in reality, but works well for this simulation (see Section \ref{section:secondary}).

\subsubsection{Outcomes}

Figures \ref{poor-result} and \ref{thor-result} show the wild tick populations in the control regions after six years of sterile release. The thoroughly distributed release has been significantly more effective: the mean population in the poor release region ($0.164$) is $62\%$ higher than that in the thorough release region ($0.102$). Ten years or $2.35$ times the release would be required for the poor release to match the thorough release. This demonstrates why low dispersal is cited as a principal obstacle to tick SIT \cite{lance2021mass}.

\section{Discussion}

\subsection{Program durations}

Recall that I sought to address the concern that \textit{I. scapularis} SIT might take too long. Selecting release rates using ideal program criteria, I found that programs would be completed in eight years or less for nearly all parameter value sets. Given that a good deal of parameter variation is to be expected within any decent-sized control area because of habitat heterogeneity \cite{tran2021spatio,gaff2020lymesim}, and that a significant fraction of cases (around $1/4$) require eight years for the safer criteria, it seems likely that some part of the control area will last this long. Thus, my model suggests that most sizable release programs will take eight years.

The best way to assess the implications of this result is to compare to SIT programs for other species that have been considered successful. The earliest SIT operation, targeting screwworm flies, took no more than a year in many regions, so my model does not suggest that ticks would be controlled outstandingly quickly \cite{novy1991screwworm}. However, some subsequent programs comparable to what I simulated have taken longer and still been deemed cost-effective, such as a fruit fly program in northern Chile that took six years \cite{enkerlin2021impact}; so \textit{I. scapularis} SIT is also not unprecedentedly slow. This is a promising result: despite \textit{I. scapularis}'s long life course, program duration may not be a dominant concern with SIT.

\subsection{Sterile tick distribution}\label{section:meta-results}

In his seminal paper on SIT, E.F. Knipling proposed five criteria for a species to be suited to SIT; among them is the capability of the released males to be rapidly dispersed to all corners of the control region (assuming the region is bounded such that wild ticks are present throughout) \cite{knipling1955possibilities}. Since \textit{I. scapularis} does not move much, this means efficient and thorough manual dispersal might be necessary for \textit{I. scapularis} SIT \cite{lance2021mass,alphey2010sterile,carroll1996dispersal}.

My results substantiate the importance of distributing the sterile ticks to all parts of the control area: a good deal more time and/or sterile ticks are required for even slightly less than thorough release. Thorough distribution certainly necessitates aerial release \cite{dowell2021supply}, which unfortunately is often a sizable portion of SIT program expense (when used), in some cases increasing costs by as much as $70\%$ \cite{esch2021operational,tan2013alternative}. This is much lower than the increase in release needed for the ``poor'' distribution, but is similar to the change in time required. However, the ``poor'' scenario was still a nearly complete distribution (Section \ref{section:meta-result}). Assuming cost scales roughly linearly with time, my model thus suggests non-aerial release must achieve nearly as complete a distribution as aerial release to be more cost-effective even than the most expensive aerial release; it is not plausible for non-aerial release to be this thorough \cite{dowell2021supply}, so aerial release is optimal despite its cost.

In light of the expected unusual expense of rearing \textit{I. scapularis}, discussed next (Section \ref{section:rearing}), the cost of aerial release is cause for serious concern. Fortunately, increased usage of aerial release for SIT has brought more attention to this method \cite{tween2007current,tan2013alternative,mubarqui2014smart,bouyer2020field}, and strides are being made to reduce its cost \cite{dowell2021supply,tween2007current,mirieri2020new,bouyer2020field}. Such progress may be a prerequisite for \textit{I. scapularis} SIT.

\subsection{Rearing\texorpdfstring{\;of\;\,}{\ of }\textit{I. scapularis}}\label{section:rearing}
A notable challenge and presumed exceptional expense in\\\textit{I. scapularis} SIT, which is beyond the scope of my model but nevertheless worth mentioning, is the rearing of large numbers of ticks. \textit{I. scapularis} rearing in the lab has already been put into practice \cite{troughton2007life,burtis2021susceptibility}, but the scale required for SIT is formidable \cite{parker2021mass}. Rearing constitutes a major expense in all SIT programs; it is the ticks' long life course that makes this of particular concern in \textit{I. scapularis} SIT. The \textit{I. scapularis} life cycle takes at least 200 days in the lab and includes two engorgements for males and three for females \cite{troughton2007life}. This sets ticks well apart from other SIT targets \cite{zethner1980five,schwarz1985mass,gooding1997care,capinera2012sweetpotato,zheng2015standard,carvalho2014mass}; screwworm flies, for instance, develop in less than a month \cite{vargas2021impact}.

The alternative of releasing ticks early may be even less feasible. Doing so would presumably subject sterilized ticks to the same mortality as wild ticks, which primarily derives from questing failure \cite{ogden2005dynamic,mount1997simulation,tufts2021ixodes}. The survival rate to mating of ticks in the wild is on the order of $1$ in $1000$-$2000$ \cite{sandberg1992comprehensive,mount1997simulation}, forcing early release to be scaled accordingly (or at least substantially, if release is still delayed some). Aside from the logistical and economic challenges, this approach would increase ethical concerns with the release not only of more ticks, but of ticks that will feed multiple times and hence can transmit diseases.

Rearing may thus constitute a principal hurdle to \textit{I. scapularis} SIT. The expense will likely necessitate cost reduction in other areas such as release, as noted in Section \ref{section:meta-results}.

\subsection{Principal simplifications of the model}

Numerous details have been omitted from my model, including tick life course specifics, seasonal dynamics (such as temperature and moisture), habitat heterogeneity, host behavior, and stochasticity \cite{randolph2013ecology,nadolny2018modelling,barclay2021mathematical}. Perturbations in results that would stem from these factors are mitigated: in program duration estimation, by focusing on the safer criteria; in poor release testing, by direct comparison to thorough release results from the same model. Certainly, performing further tests with sterile release incorporated into more detailed models of \textit{I. scapularis}, such as LYMESIM \cite{gaff2020lymesim}, would be instructive.

\subsection{Conclusion}

I initially identified two key testable concerns with \textit{I. scapularis} SIT: that the ticks' long life course could make SIT take too long and that low migration might mean sterile males need thorough manual dispersal. To test these concerns, I developed a deterministic, discrete-time population model of \textit{I. scapularis} SIT. The model predicts that program durations will not be far from the normal range, which bodes well for \textit{I. scapularis} SIT. However, the model demonstrates that thorough manual sterile tick dispersal is indeed essential, which will likely raise expense substantially until aerial release becomes more cost-effective. With unusually high rearing expenses also expected for \textit{I. scapularis}, aerial release cost reduction may be a prerequisite to \textit{I. scapularis} SIT.

Whether \textit{I. scapularis} SIT effectiveness can match up to its costs remains to be assessed \cite{osburn1982potential}. Next steps in studying this control method for \textit{I. scapularis} could include increasingly detailed and cost-oriented modeling, additional biological research of \textit{I. scapularis} sterilization and sexing, and field testing \cite{barclay2021mathematical,thome2010optimal,kocan2015insights,benedict2021sterile}.

\newpage
\appendix
\section{Algebra}\label{appendix:algebra}

\subsection{Rearranging the no-migration model}\label{appendix:a1}

Substituting Equations \ref{W} and \ref{X} into Equation \ref{Allee} yields:
\begin{align*}
N_{t+1} &= N_t \cdot \frac{\lambda\cdot\frac{\frac{1}{2}N_t}{\frac{1}{2}N_t+S} \frac{\frac{1}{2}N_t+S}{\frac{1}{2}N_t+A+S}}{1+\left(\lambda\cdot\frac{\frac{1}{2}N_t}{\frac{1}{2}N_t+S} \frac{\frac{1}{2}N_t+S}{\frac{1}{2}N_t+A+S}-1\right) N_t} \\[5pt]
&= N_t \cdot \frac{\lambda\cdot\frac{\frac{1}{2}N_t}{\frac{1}{2}N_t+A+S}}{1+\lambda\cdot\frac{\frac{1}{2}N_t^2}{\frac{1}{2}N_t+A+S}-N_t} \\[5pt]
&= N_t \cdot \frac{\frac{1}{2}\lambda N_t}{\frac{1}{2}\lambda N_t^2 + (1-N_t)\left(\frac{1}{2}N_t+A+S\right)} \\[5pt]
&= N_t \cdot \frac{1}{N_t+\frac{1-N_t}{\frac{1}{2}\lambda N_t}\left(\frac{1}{2}N_t+A+S\right)}.
\end{align*}

\subsection{Solving for equilibria}\label{appendix:a2}

Following Appendix \ref{appendix:a1}, equilibria of Equation \ref{final} (denoted $N^*$), occur at $N^*=0$ or:
\begin{gather*}
1 = \frac{\frac{1}{2}\lambda N^*}{\frac{1}{2}\lambda (N^*)^2 + (1-N^*)\left(\frac{1}{2}N^*+A+S\right)} \\
\Rightarrow \\
\frac{1}{2}\lambda (N^*)^2 + (1-N^*)\left(\frac{1}{2}N^*+A+S\right) = \frac{1}{2}\lambda N^*.
\end{gather*}

This quadratic equation has roots at $N^*=1$ and $N^*=\frac{A+S}{\frac{1}{2}(\lambda-1)}$.

\subsection{Computing the Allee strength (\texorpdfstring{$A$}{A})}\label{appendix:a3}

The Allee threshold ($T$) is simply the nonconstant equilibrium found above and shown in Figure \ref{bifurc}, when $S=0$:
\begin{equation*}
T = \frac{A}{\frac{1}{2}(\lambda-1)}.
\end{equation*}

Rearranging yields:
\begin{equation*}
A = \frac{1}{2}(\lambda-1)T.
\end{equation*}

\newpage
\bibliographystyle{ieeetr}
\bibliography{Current}

\end{document}